\documentclass[11pt, a4paper]{article}
\usepackage[centertags,reqno]{amsmath}
\usepackage{amssymb}
\usepackage{fancyhdr}
\usepackage{microtype}
\usepackage[hmargin=3.3cm,vmargin=3.5cm]{geometry}
\usepackage{graphicx}

\usepackage[breaklinks]{hyperref}

\hypersetup{
    pdftitle={Lambda: A Mathematica-package for operator product expansions in vertex algebras},    
    pdfauthor={ Joel Ekstrand},     
    pdfnewwindow=true,      
    colorlinks=true,       
    linkcolor=black,          
    citecolor= black,        
    filecolor= black,      
    urlcolor= black           
}

\numberwithin{equation}{section}

\newcommand{\ie}{{\it i.e.}}
\newcommand{\eg}{{\it e.g.}}
\newcommand{\uD}{D}
\newcommand{\ud}{d}

\newcommand{\LB}[3][\Lambda]{[\, #2  \,_{#1}\,  #3 \,]}
\newcommand{\lB}[3][\lambda]{[\, #2  \,_{#1}\,  #3 \,]}

\newcommand{\cmd}[1]{\texttt{#1}}

\begin{document}
\title{Lambda: A Mathematica-package for operator product expansions in vertex algebras}
\date{2010 }
\author{Joel Ekstrand\footnote{Email address: \href{mailto:joel.ekstrand@physics.uu.se}{joel.ekstrand@physics.uu.se}} \\ Department of Physics and Astronomy, Uppsala University}
\maketitle
\begin{abstract}
\noindent We give an introduction to the Mathematica package Lambda, designed for calculating $\lambda$-brackets in both vertex algebras, and in SUSY vertex algebras. This is equivalent to calculating operator product expansions in  two-dimensional conformal field theory. The syntax of $\lambda$-brackets is reviewed, and some simple examples are shown, both in component notation, and in $N=1$ superfield notation.
     \end{abstract}
     
\vspace{3mm}
\noindent\textbf{Note added (2026):} The package is now maintained on GitHub: \url{https://github.com/joellernej/lambda}. An issue regarding the normalization of the odd integration variable in the SUSY right quasi-Leibniz rule has been corrected (thanks to N.~Tanabe for pointing out the issue and providing the fix). The author can be reached at \href{mailto:joel@fejston.se}{joel@fejston.se}.
\vspace{1mm}

\thispagestyle{fancy}

\section{Introduction}
Conformal field theory is one of the cornerstones of theoretical physics.
Vertex algebras, originally introduced in \cite{Borcherds:1983sq}, provide a rigorous mathematical formulation of the chiral part of two-dimensional conformal field theory. The axioms of a vertex algebra are obtained from an abstract treatment of the properties of quantum field theories, and of   \textit{operator product expansions} (OPEs).

The key  idea of OPEs is that a product of local operators  defined at nearby locations can be expanded in a series of local operators.
The formalism developed within vertex algebras provides a very compact and computational convenient way for  calculating OPEs, including quantum effects.
The reference to the position of the operators is abstracted away, and the OPEs are encoded in so-called $\lambda$-brackets~\cite{DAndrea:1998p403}.
There is also a normal order product, capturing the regular part of the product of two operators.
 The relations between these two products, in other words, formulas of OPEs between composite fields, are algebraic. The calculations are not necessarily  hard to perform by hand, but they rapidly grow in number, and thus become error-prone. This makes them very appropriate for being implemented in some computational software, like Mathematica.
This is the purpose of the Mathematica package at hand: Lambda. 

The package grew out of a need to calculate $\lambda$-brackets between composed operators, and the need to do so in $N=1$ superfield formalism. This was motivated by the construction of the \textit{Chiral de Rham complex} (CDR), introduced in \cite{Malikov:1998dw}.  The CDR is a sheaf of vertex algebras,  and investigation of symmetries of this sheaf relies on the formalism of $\lambda$-brackets, see, for example, \cite{benzvi-2006} and \cite{heluani-2008}.  In \cite{Ekstrand:2009zd}, the CDR was given the interpretation of a formal Hamiltonian quantization of the supersymmetric non-linear sigma model. Thus, insights from the sigma model could be used to further investigate symmetries of the CDR. In  \cite{Ekstrand:2010wu}, the symmetries of the CDR on special holonomy manifolds were investigated. Here, the package Lambda was heavily used, and the OPEs between operators composed of up to five fields were calculated.

There exists other implementations in Mathematica, and similar computational software, for calculating OPEs. The most used seems to be OPEdefs~\cite{Thielemans:1991uw}. Also see~\cite{Fujitsu:1994ae}.  The package OPEdefs  has been developed further to handle $N=2$ superfield in~\cite{Krivonos:1995bk}. One main advantage of the package Lambda is that it implements the algebraic approach to OPEs given by vertex algebras. It is the first package to implement the efficient $\lambda$-brackets. It also handles indices, and operators can be constructed out of tensors. Both input and output are handled graphically, thus increasing the readability in the usage of the module. It is also tested in Mathematica 7, and therefore offers a newer environment than the previous packages.

The paper is organized as follows. 
In section \ref{sec:OPE}, the basic properties of vertex algebras are reviewed, along with the relations between OPEs and $\lambda$-brackets. The aim is to give an understanding of how to interpret the results of the package Lambda in the, perhaps more familiar, language of OPEs.  In section \ref{sec:SOPE}, this is generalized to $N=1$ superfields. Section \ref{LBrules} lists the basic relations between the different operators and brackets. In section \ref{sect:howtouse}, we give some examples of how to use the package Lambda. Section \ref{sect:listofcommands} lists the commands provided by Lambda.

\subsection*{Acquiring the package}
The package is available from the Computer Physics Communications library, see \url{http://cpc.cs.qub.ac.uk/}. It can also be downloaded from \url{http://www.fejston.se/lambda}.

\section{Operator product expansions and vertex algebras}\label{sec:OPE}

In operator product expansions, the product of two operators at nearby points  is expressed as a sum of operators: $A_i(z) A_j(w) = \sum_k f^k_{ij} (z-w) A_k(w) $,
where each operator $A_k$ is well behaved at $w$, and $f^k_{ij} (z-w)$ is a function, possible divergent when $z \rightarrow w$. 
In two dimensions, the positions are coordinates on the complex plane, and the OPE is a Laurant series.
Writing out the poles, and only keeping divergent terms, we can write the OPE between the operators $A$ and $B$ as
\begin{equation} \label{eq:OPEexample}
A(z) B(w) \sim \frac{(A_{(0)}B)(w)}{(z-w)} + \frac{(A_{(1)}B)(w)}{(z-w)^2} + \frac{(A_{(2)}B)(w)}{(z-w)^3} + \ldots ~,
\end{equation}
where $(A_{(j)}B)$ are operators.
Two operators are said to be mutually local if  there is a highest pole, \ie, if the series \eqref{eq:OPEexample} terminates. 

A vertex algebra captures the structure of such OPEs. We here give a brief overview of the concepts of a vertex algebra, emphasizing the similarities with OPEs. For a proper introduction to vertex algebras, see \cite{Kac:1996wd} and \cite{Frenkel:2004jn}. 

A vertex algebra is a vector space $V$ (the \textit{space of states}), 
with a vector $|0\rangle \in V$ (the \textit{vacuum}), 
a map  $Y$ from a given state $A \in V$ to a field $Y(A,z)$ (called  the  \textit{state-field correspondence}), 
and  an endomorphism $\partial: V\rightarrow V$ (the \textit{translation operator}). 

A \textit{field} is defined as an $\text{End}(V)$-valued distribution in a formal parameter $z$:
\begin{equation}
A(z)=\sum_{j\in\mathbb{Z}}  \frac{1}{z^{j+1} } A_{(j)},\quad \text{where } A_{(j)}\in \text{End}(V) ~, 
\end{equation}
and for all $B \in V$, $A(z)B$ contains only finitely many negative powers of $z$.  The field $Y(A,z)$ will also be denoted by $A(z)$.

These structures are subject to certain axioms. For instance, the vacuum should be invariant under translations:  $\partial |0\rangle = 0$. Acting with $\partial$ on the endomorphisms of a field, should be the same as differentiation of the field with respect to the formal parameter~$z$:
\begin{equation}
 {[}\partial, Y(A,z)] = \partial_z Y(A,z)~.
\end{equation}
We will use $\partial$ to denote both the endomorphism and $\partial_z$.
The state-field correspondence creates a given state $A$ from the vacuum in the limit $z \rightarrow 0$: 
\begin{equation}
Y(A,z) |0\rangle |_{z=0} = A_{(-1)}|0\rangle = A ~.
\end{equation}

From the endomorphisms $A_{(j)}$ of $Y(A,z)$ (called the \textit{Fourier modes}), we can define the \textit{$\lambda$-bracket}:
\begin{equation} \label{eq:lambdabracketdef}
\lB{A}{B}  =  \sum_{j \geq 0}   \frac{\lambda^j}{j!}  (A_{(j)}B)   ~.
\end{equation}
The $\lambda$-bracket can be viewed as a formal Fourier transformation of $Y(A,z)B$:
\begin{equation} 
\lB{A}{B} =  \text{Res}_z \; e^{\lambda z} \; Y(A,z)B ~,
\end{equation}
where $\text{Res}_z$ picks the $z^{-1}$-part of the expression. 
The locality axiom of the vertex algebra says that the series \eqref{eq:lambdabracketdef} terminates for all $A$ and $B$, in other words, all fields in a vertex algebra are mutually local.

The $\lambda$-bracket \eqref{eq:lambdabracketdef} encodes the same information as the OPE \eqref{eq:OPEexample}. For example, the operator expansion of a Virasoro generator $L$ is
\begin{equation} \label{eq:opeVir}
L(z) L(w) \sim \frac{1}{(z-w)} \partial L(w) + \frac{2}{(z-w)^2} L(w) + \frac{c}{2(z-w)^4} ~,
\end{equation}
where $c$ is the central charge. In $\lambda$-bracket notation, this is written
\begin{equation}\label{eq:lambdabrLL}
\lB{L}{L}= ( \partial + 2 \lambda ) L + \frac{c}{12} \lambda^3 ~.
\end{equation}

One way to define a normal order product of operators in quantum field theory is by point splitting. The singular terms captured by the OPE between the operators, are subtracted from the product:
\begin{equation}
(A B)(z) = \lim_{w\rightarrow z}\left ( A(w) B(z) - \text{singular terms} \right )~,
\end{equation}
and the result is a well-defined operator at $z$.
In vertex algebras, the  \textit{normal ordering product} is defined in a similar way; by acting with the non-singular part of a field $A(z)$, in the limit $z \rightarrow 0$:
\begin{equation}
: A B : \; = A_{(-1)} B~.
\end{equation}
It is important to stress that this product is neither commutative, nor associative, see \eqref{eq:Quasicommutativity} and \eqref{eq:Quasiassociativity}. This is a quantum effect.
In all further considerations, we drop the notation $:\;\;:$, and the normal ordering is always assumed in products.

The above construction easily extends to the case when $V$ is a super vector space. The state-field correspondence $Y$ should respect this grading, $\partial$ should be an even endomorphism, and the vacuum should be even. 

For instance, in a Neveu-Schwarz superconformal algebra, there is,  in addition to $L$ with the OPE \eqref{eq:opeVir}, an odd operator $G$, with the OPEs
\begin{subequations}  \label{eq:opeSuperConf}
\begin{align}
G(z) G(w) &\sim \frac{2}{(z-w)} L(w) + \frac{2 c}{3 (z-w)^3} ~, \\
L(z) G(w) &\sim    \frac{3}{2 (z-w)^2} G(w) + \frac{1}{(z-w)} \partial G(w) ~. \label{eq:opeSuperConfLG}
\end{align}
\end{subequations}
In the language of $\lambda$-brackets, this is equivalent to
\begin{subequations}  \label{eq:lambdabrN1}
\begin{align}
\lB{G}{G} &= 2 L + \tfrac{c}{3} \lambda^2 ~, \\
\lB{L}{G} &= \left( \partial + \tfrac{3}{2} \lambda \right) G~.
\end{align}
\end{subequations}

The relations and properties of the introduced operators $\partial$, $\lB{\cdot}{\cdot}$ and the normal order product will be listed in section \ref{LBrules}. Before that, we introduce a superfield formulation of OPEs and of vertex algebras. 

\section{OPEs of superfields and SUSY vertex algebras}\label{sec:SOPE}
Combining  conformal field theory with supersymmetry, one gets superconformal field theory. This can be formulated in terms of $N=1$ superfields. Adding a fermionic (odd) partner $\theta$ to the bosonic (even) coordinate $z$, allows us to combine two fields  of opposite grading, $A(z)$ and $B(z)$, into a superfield: $C(z,\theta) = A(z) + \theta B(z)$. To describe the operator product expansions of superfields, we let $Z_i=(z_i, \theta_i)$, and define the displacements
\begin{align}
Z_{12} &= z_1 - z_2 - \theta_1 \theta_2 ~, & \theta_{12} = \theta_1 - \theta_2 ~.
\end{align}
and the derivative $D_i = \partial_{\theta_i} + \theta_i \partial_{z_i}$
An OPE between superfields can now be written as
\begin{equation} \label{eq:SuperOPE}
\begin{split}
A(Z_1) B(Z_2) \sim & \frac{   \theta_{12}   (A_{(0|0)} B)(Z_2) }{Z_{12} } +\frac{  (A_{(0|1)} B)(Z_2) }{Z_{12} } + \frac{   \theta_{12}   (A_{(1|0)} B)(Z_2) }{Z_{12}^2 } + \\
& +\frac{  (A_{(1|1)} B)(Z_2) }{Z_{12}^2 } + \frac{   \theta_{12}   (A_{(2|0)} B)(Z_2) }{Z_{12}^3 } +\frac{  (A_{(2|1)} B)(Z_2) }{Z_{12}^3 } + \ldots
\end{split}
\end{equation}
where $(A_{(j|J)} B)$ are operators.
Note that the poles are expanded as
\begin{align}
\frac{1}{Z_{12}^{n}} &= \frac{1}{(z_1 - z_2)^n} + \frac{ n \theta_1 \theta_2}{(z_1 - z_2)^{n+1}} ~, & \frac{\theta_{12}}{Z_{12}^{n}} &= \frac{\theta_{12}}{(z_1 - z_2)^n} ~,
\end{align}
so, \eg, $Z_{12}^{-1}$ contains a double pole in $z_1-z_2$.

We exemplify this by  combining the operators $L$ and $G$ of the superconformal algebra \eqref{eq:opeSuperConf}, into one odd superfield operator:
\begin{equation}
P(z,\theta) = G(z) + 2 \theta L(z) ~.
\end{equation}
To calculate the OPE of $P$ with itself, we need \eqref{eq:opeVir} and \eqref{eq:opeSuperConf}. The OPE $G(z)L(w)$ can be calculated from \eqref{eq:opeSuperConfLG} by expanding $G(z)$ around $G(w)$: 
\begin{equation}
G(z) L(w) \sim   \frac{3}{2 (z-w)^2} G(w) + \frac{1}{2 (z-w)} \partial G(w)~.
\end{equation}
Using this, the superconformal algebra is given by 
\begin{equation} \label{eq:OPEPP}
P(Z_1) P(Z_2) \sim  \frac{\theta_{12}  }{Z_{12}} 2 \partial_2 P(Z_2) +  \frac{1 }{Z_{12}} D_2 P(Z_2)   +   \frac{\theta_{12} }{Z_{12}^2} 3 P(Z_2)   +  \frac{2 c }{3 Z_{12}^3}  ~.
\end{equation}

An efficient treatment of the OPEs of $N=1$ superfields is given by the $N_K=1$ SUSY vertex algebra~\cite{Heluani:2006pk}, which extends the vertex algebra construction of the last section. For details, see \cite{Heluani:2006pk}. For a similar superfield treatment of vertex algebras, see \cite{1996q.alg}.
Let  $\theta$  be an odd formal parameter, with $\theta^2 = 0$. Define superfields by 
\begin{equation}
A(z,\theta)=\sum_{j\in\mathbb{Z}}  \frac{1}{z^{j+1}} \left( A_{(j|1)} +  \theta \; A_{(j|0)} \right ), \quad A_{(j|J)}\in \text{End}(V) ~.
\end{equation}
In a SUSY vertex algebra, the state-field correspondence maps a state $A$ to a superfield: $Y(A,z,\theta) = A(z,\theta)$. We have an odd translation operator $D$, that acts on a field by
\begin{equation}
Y( D A, z, \theta ) = D \; Y(  A, z, \theta ) = (\partial_\theta + \theta \partial_z)  A( z, \theta ) ~,
\end{equation}
where we again use the same symbol for the endomorphism, and the action on a field. The square of this operator is the even translation operator: $D^2 = \partial$.

The action of the Fourier modes of a superfield $A(z,\theta)$ on another field is described by the $\Lambda$-bracket:
 \begin{equation} \label{eq:Lambdabracket}
            \LB{A}{B} = \sum_{j \geq 0}
            \frac{\lambda^j}{j!} \left ( A_{(j|0)}B + \chi \; A_{(j|1)}B \right ) ~, 
    \end{equation}
     where $\lambda$ is an even, and  $\chi$ an odd parameter,  satisfying $\chi^2 = - \lambda$. We write  $\LB{\cdot}{\cdot}$, instead of $\lB{\cdot}{\cdot}$, to indicate that we work with superfields.
The normal order product is given by
 \begin{equation}
  A B = A_{(-1|1)}B ~,
    \end{equation}
and again,  this product is neither commutative, nor associative, see \eqref{eq:Quasicommutativity} and \eqref{eq:Quasiassociativity}.

The $\Lambda$-bracket \eqref{eq:Lambdabracket} contains the same information as the OPE \eqref{eq:SuperOPE}. For instance, 
the OPE \eqref{eq:OPEPP} is written
\begin{equation} \label{eq:LambdabracketPP}
\LB{P}{P} = (2 \partial + \chi D + 3 \lambda ) P + \frac{c}{3} \lambda^2 \chi ~.
\end{equation}

We now review some relations between the introduced operations.

\section{\texorpdfstring{$\Lambda $}{Lambda}-bracket calculus}\label{LBrules}

In this section we collect some properties of $\Lambda$-bracket calculus. For further explanations 
 and details, the reader may consult \cite{Kac:1996wd} and \cite{Heluani:2006pk}. We give a unified description of the vertex algebras and the SUSY vertex algebras. Let $N$ denote whether we are working with superfields ($N=1$) or not  ($N=0$).
The brackets $\LB{\cdot}{\cdot}$ below are to be interpreted as $\Lambda$-brackets ($N=1$) or $\lambda$-brackets ($N=0$), depending on in which context they are used.  We use the same symbol for an operator and its grading, the meaning should be clear. The relevant commands provided by Lambda are listed (in \cmd{monospace typewriter} font). For further explanations of the commands, see section~\ref{sect:listofcommands}.
    \begin{itemize}
        \item Relations between the translation operators and  $\lambda$ and $\chi$:
        		\begin{align}
\uD ^2&=\partial & [\uD,\partial]&=0 & [\uD,\lambda]&=0  & [\partial,\lambda]&=0  \\
\chi^2&=-\lambda 	 & [\uD,\chi]&=2\lambda  & [\partial,\chi]&=0  
\end{align}
        \item Sesquilinearity:
            \begin{align}
\LB{\uD a}{ b} &=  \chi \LB{a}{b}
& \LB{a}{ \uD b} &= -(-1)^{a} \left( \uD
                + \chi
                \right) \LB{a}{b}
\\
\LB{\partial a }{ b} &=  -\lambda \LB{a}{b}
&\LB{a}{\partial  b} &=  \left( \partial 
               	+ \lambda
               	\right) \LB{a}{b}   
            \end{align}
        \item Skew-symmetry:
            \begin{equation}                 \label{eq:Skewsymmetry}
                \LB{a}{b} =  - (-1)^{a b + N} \LB[-\Lambda -
                \nabla]{b}{ a}
            \end{equation}
            The bracket on the right-hand side is in the SUSY case computed by first computing $\LB[\Gamma]{b}{a}$, where $\Gamma =
            (\rho, \eta)$, $\rho$ even and $\eta$ odd. Then $\Gamma$ is replaced by $(-\lambda - \partial,
            -\chi - \uD)$. In the $N=0$ case, compute $\lB[\rho]{b}{a}$, and replace $\rho$ by $-\lambda - \partial$. This rule is implemented by the function \cmd{LambdaBracketChangeOrder[a,b]} .
        \item Jacobi identity:
            \begin{equation}\label{eq:Jacobiid}
               \LB{a}{\LB[\Gamma]{b}{c}} = (-1)^{N(a+1)} \LB[\Gamma + \Lambda]{ \LB{a}{b}  }{ c} +
                (-1)^{(a+N)(b+N)} \LB[\Gamma]{b}{ \LB{a}{c} }
            \end{equation}
            where the first bracket on the right hand side is computed as in \eqref{eq:Skewsymmetry}.
        \item Quasi-commutativity:
            \begin{equation}\label{eq:Quasicommutativity}
                ab - (-1)^{ab} ba = \int_{0}^\nabla \LB{a}{b} \ud\Lambda
            \end{equation}
            The integral $\int d\Lambda$ is $\int d\lambda$ in the $N=0$ case,  and $\partial_\chi \int d\lambda$ in the SUSY case. The limits of the integral mean that $\lambda$ should be replaced by $\partial$. Implemented by \cmd{NormalOrderChangeOrder[a,b]} .
        \item Quasi-associativity:
            \begin{equation}   \label{eq:Quasiassociativity}
             (ab)c - a(bc) =   \left( \int_0^{\nabla} d\Lambda a \right)  \LB{b}{c} + (-1)^{a b} \left( \int_0^{\nabla} d\Lambda b \right)  \LB{a}{c}
            \end{equation}
            This formula is to be understood as follows. First, the $\Lambda$-brackets are calculated. The $\lambda$'s and $\chi$'s are integrated as in \eqref{eq:Quasicommutativity}. The resulting operators (\ie,  a factor times $\partial$ to some power) act on $a$ respectively $b$, and this is normal ordered with the resulting operators from the brackets. This is implemented in \cmd{Normal\-Order\-Change\-Parenthesis[expr]}.
        \item Quasi-Leibniz (non-commutative Wick formula):
            \begin{equation}
                \LB{a}{b c }  = \LB{a}{b} c + (-1)^{(a+N)b}b
               \LB{ a}{c} + \int_0^\Lambda \LB[\Gamma]{ \LB{a}{b} }{c} \ud \Gamma
            \end{equation}
            The integration is to be understood as in \eqref{eq:Skewsymmetry} and \eqref{eq:Quasicommutativity}. The limits of the integral mean replacing $\rho$ by $\lambda$. 
\end{itemize}

\section{Example of how to use Lambda} \label{sect:howtouse}
In this section we give some examples of how to use Lambda.
We demonstrate a realization of the $N=1$ superconformal algebra.
We first give the algebra in components, \ie, without using superfields. After this, the algebra is written in $N=1$ superfield formalism. We then show how to extend this to $N=2$ superconformal algebra, by an operator constructed out of a $2$-form. 
The given examples are quite simple, but hopefully still illuminating. It is, in general, however, possible to define algebras with arbitrary complexity, including non-linear ones, by giving consistent definitions of the function \cmd{Lambda\-Bracket}.

First, we must make sure that the package is in a directory where Mathematica can find it, for example by adding \cmd{AppendTo[\$Path, "pathtolambda"]} in the initialization file \cmd{\$UserBaseDirectory/Kernel/init.m}, or by using the command \cmd{SetDirectory}.  After this, we can load the package Lambda:
\begin{flushleft}
\hspace{1mm}\includegraphics[scale= 0.73]{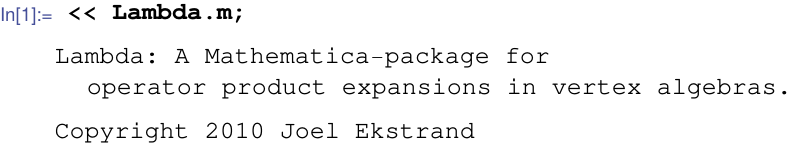}
\end{flushleft}
\subsection{$N=1$ superconformal algebra in components}
We start without superfields, and write
\begin{flushleft}
\hspace{1mm}\includegraphics[scale= 0.73]{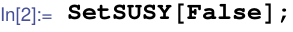}
\end{flushleft}
Let $\gamma^\mu$ and $\beta_\nu$ be even fields. We define this with the following code:
\begin{flushleft}
\hspace{1mm}\includegraphics[scale= 0.73]{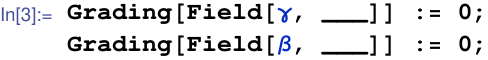}
\end{flushleft}
A field is written with the command \cmd{Field}. The first argument is the symbol of the field, the second is a list of the lower indices that the field carries, the third is the upper indices, and the fourth can be used as target space derivatives on the field. A field is denoted with a hat over it, and the package recognizes any symbol with a hat as a field. As an example, consider the field 
$\gamma^i$. It has one upper index, and can be written as
\begin{flushleft}
\includegraphics[scale= 0.73]{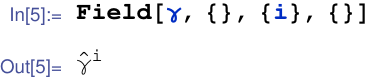}
\end{flushleft}
Mathematica also recognizes this form as input. 
We want our fields to have the brackets 
 \begin{align}
\lB{\beta_i}{\gamma^j} &= \hbar \delta_i^j~, & \lB{\beta_i}{\beta_j} &= 0~, & \lB{\gamma^i}{\gamma^j} &=0 ~.
\end{align}
In order for the package to know that the constant $\hbar$ is not an expression that potentially contains a field, it needs to be declared to be  \cmd{NumericQ}:
\begin{flushleft}
\hspace{1mm}\includegraphics[scale= 0.73]{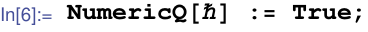}
\end{flushleft}
There is a predefined even Kronecker delta field with symbol $\delta$ in the package. The bracket $\lB{\beta_i}{\gamma^j}=\hbar \delta_i^j$ is then specified by
\begin{flushleft}
\hspace{1mm}\includegraphics[scale= 0.73]{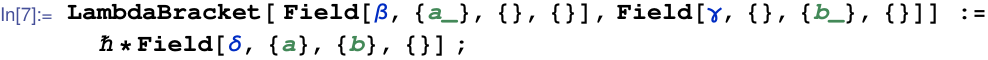}
\end{flushleft}
This defines the bracket with the arguments in this specific order. To define the bracket in the opposite order, we use the function \cmd{Lambda\-Bracket\-Change\-Order}, that implements the skew-symmetry property  \eqref{eq:Skewsymmetry}:
\begin{flushleft}
\hspace{1mm}\includegraphics[scale= 0.73]{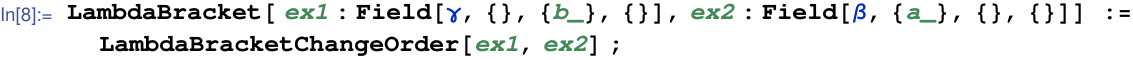}
\end{flushleft}
Note that it is possible to define the opposite order bracket without using  \cmd{Lambda\-Bracket\-Change\-Order}. Sometimes this can be useful, if one wish to have a specific form of the output (other than the one given by \cmd{Lambda\-Bracket\-Change\-Order}). Care has then to be taken so that the defined bracket fulfills skew-symmetry \eqref{eq:Skewsymmetry}.

The other two brackets are given by
\begin{flushleft}
\hspace{1mm}\includegraphics[scale= 0.73]{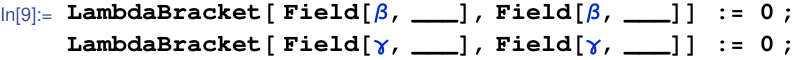}
\end{flushleft}
Now, these fields are a so-called  $\beta \gamma$-system. 
We can define a Virasoro field $L_{\text{bos}}=\beta_i \partial \gamma^i$ by
\begin{flushleft}
\hspace{1mm}\includegraphics[scale= 0.73]{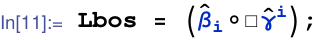}
\end{flushleft}
The circle denotes normal ordering, and the square\footnote{The reason to use a square is that it is an operator without a built in meaning in Mathematica, in contrast to the symbol  $\partial$.}  is the (even) derivative $\partial$.  The bracket of this field with itself is calculated by
\begin{flushleft}
\includegraphics[scale= 0.73]{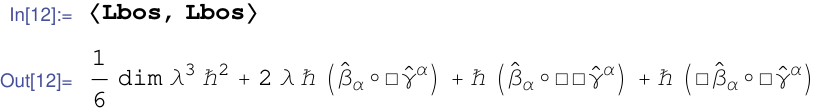}
\end{flushleft}
The derivatives of this expression can be collected using \cmd{CD} (an alias for \cmd{Collect\-Derivatives}):
\begin{flushleft}
\includegraphics[scale= 0.73]{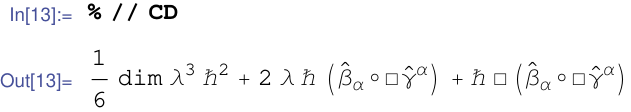}
\end{flushleft}
This shows that $L_{\text{bos}} /\hbar$ satisfies the Virasoro algebra \eqref{eq:lambdabrLL}, with central charge $c=2$ dim, \ie, 
\begin{equation}
\lB{ L_{\text{bos}}}{ L_{\text{bos}}} = \hbar (2 \lambda + \partial ) L_{\text{bos}} + \hbar^2  \lambda^3 \frac{\text{dim}}{6} .
\end{equation}
The variable dim is equal to the trace of the Kronecker delta: dim$=\delta^a_ a$.

We now define two odd fields, $b_i$ and $c^j$, with the brackets
 \begin{align}
\lB{b_i}{c^j} &= \hbar \delta_i^j~, & \lB{b_i}{b_j} &= 0~, & \lB{c^i}{c^j} &=0 ~.
\end{align}
This is a  $\beta\gamma b c$-system. 
This is implemented in a similar way as above. The basic fields, $\beta$, $\gamma$, $b$ and $c$,  are predefined, along with their brackets,  in the file \cmd{basicfields.m} (included in the distribution of Lambda), which we now load:
\begin{flushleft}
\hspace{1mm}\includegraphics[scale= 0.73]{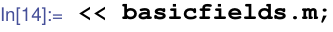}
\end{flushleft}
Of course, this file could have been loaded directly, and the above definitions are only included in order to show the basic syntax.
To make the calculations less cluttered, we set $\hbar=1$.
\begin{flushleft}
\hspace{1mm}\includegraphics[scale= 0.73]{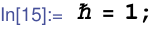}
\end{flushleft}

We can define another Virasoro field $L_\text{ferm}=\frac{1}{2}(\partial b_i c^i +\partial c^i b_ i )$ by
\begin{flushleft}
\hspace{1mm}\includegraphics[scale= 0.73]{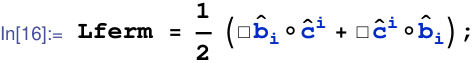}
\end{flushleft}
The bracket is
\begin{flushleft}
\includegraphics[scale= 0.73]{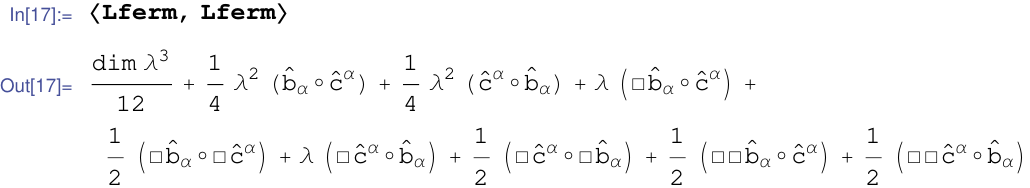}
\end{flushleft}
In order to read off the algebra, some reordering is needed. Let us change the order in the second term using \cmd{NormalOrderChangeOrder} (with alias \cmd{NOCO}) , and collect the derivatives with \cmd{CD}: 
\begin{flushleft}
\includegraphics[scale= 0.73]{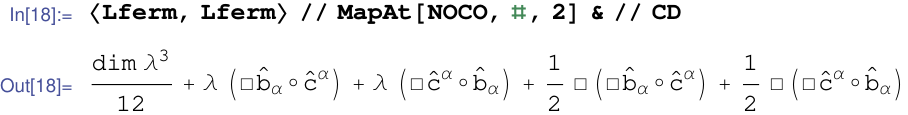}
\end{flushleft}
The Virasoro algebra \eqref{eq:lambdabrLL} is now apparent, and the central charge is $c=$dim.
Since the two Virasoro fields commute:
\begin{flushleft}
\includegraphics[scale= 0.73]{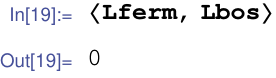}
\end{flushleft}
they can be combined  into one Virasoro field $L$:
\begin{flushleft}
\hspace{1mm}\includegraphics[scale= 0.73]{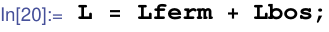}
\end{flushleft}
The field $L$ fulfills the Virasoro algebra with central charge $c=3$ dim. Let us define an odd field $G$ by
\begin{flushleft}
\hspace{1mm}\includegraphics[scale= 0.73]{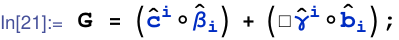}
\end{flushleft}
This field has conformal weight $\frac{3}{2}$ with respect to $L$
\begin{flushleft}
\includegraphics[scale= 0.73]{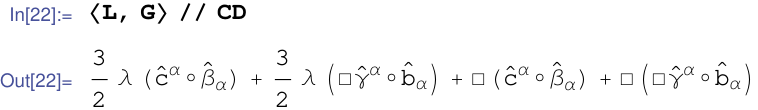}
\end{flushleft}
and $\lB{G}{G}$ is
\begin{flushleft}
\includegraphics[scale= 0.73]{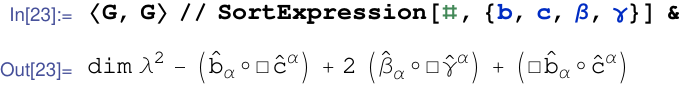}
\end{flushleft}
where we have sorted the expression. Let us compare this with
$2 L+  \lambda^2 \text{dim}$.
We sort the expression, and rename the dummy indices in a canonical way, using \cmd{Make\-Dummy\-Indicies\-Canonical} (with alias \cmd{MDIC}):
\begin{flushleft}
\includegraphics[scale= 0.73]{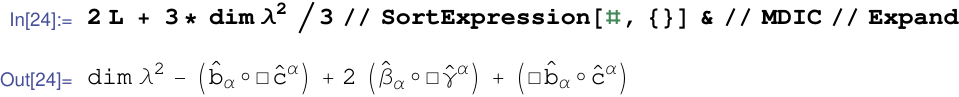}
\end{flushleft}
We see that we have the $N = 1$ superconformal algebra, given by \eqref{eq:lambdabrLL} and \eqref{eq:lambdabrN1}.
\subsection{$N=1$ superconformal algebra in superfields}
We now want to write the $N = 1$ superconformal algebra in superfields, and write 
\begin{flushleft}
\hspace{1mm}\includegraphics[scale= 0.73]{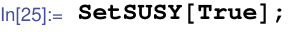}
\end{flushleft}
Let us combine the  $\beta \gamma b c$-system into two superfields. One even, $\Phi^i =\gamma^i + \theta c^i$, and one odd, $S_i= b_i + \theta \beta_i$, with the brackets
 \begin{align}
\LB{S_i}{\Phi^j} &= \delta_i^j~, & \LB{S_i}{S_j} &= 0~, & \LB{\Phi ^i}{\Phi ^j} &=0 ~.
\end{align}
The brackets and the gradings are defined as for the $\beta\gamma$-system, with the only difference that $S$ is declared to have odd grading. All the definitions are in the file \cmd{basicfields.m}.
The components of the superfields are named using \cmd{DefineComponents}:
\begin{flushleft}
\hspace{1mm}\includegraphics[scale= 0.73]{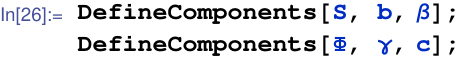}
\end{flushleft}

The Virasoro field and the supercurrent are now given by $P = \partial \Phi^i S_i + D \Phi^i D S_i$ :
\begin{flushleft}
\hspace{1mm}\includegraphics[scale= 0.73]{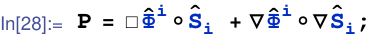}
\end{flushleft}
Here, the $\nabla$-symbol represents the odd derivative $D$. The components of $P$ are $P= G + \theta 2 L$, which can be seen by using \cmd{ExpandSuperFields}. This returns a list, with the components, and their grading:
\begin{flushleft}
\includegraphics[scale= 0.73]{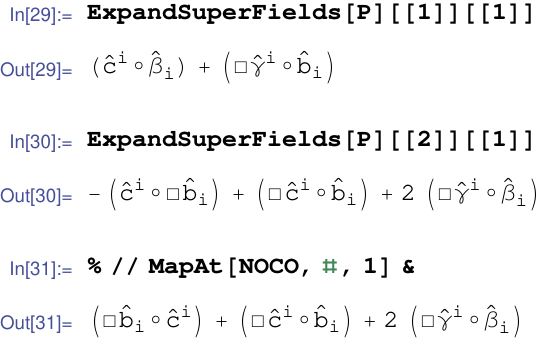}
\end{flushleft}
As expected, the field $P$ fulfills the $N=1$ superconformal algebra \eqref{eq:LambdabracketPP}, with central charge $c=3$ dim, \eg, 
\begin{equation}
\LB{P}{P} = (2 \partial + \chi D + 3 \lambda ) P + \text{dim} \lambda^2 \chi ~.
\end{equation}
This is calculated by
\begin{flushleft}
\includegraphics[scale= 0.73]{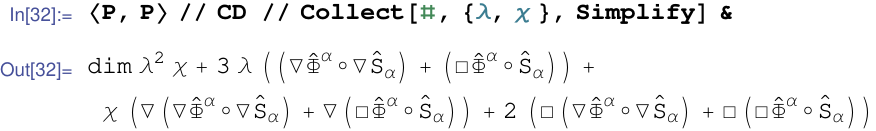}
\end{flushleft}

\subsection{$N=2$ superconformal algebra}
Now, we follow \cite{heluani-2008}, and extend this algebra to an $N=2$ superconformal algebra, by introducing a constant 2-form $\omega$, and its inverse. We define the fields by
\begin{flushleft}
\hspace{1mm}\includegraphics[scale= 0.73]{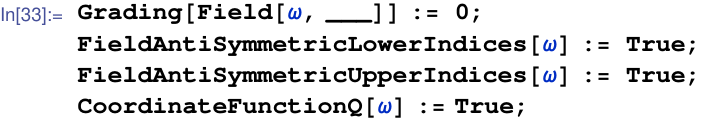}
\end{flushleft}
We use the same symbol for its inverse, with $\omega_{i j} \omega^{j k} = \delta_i^k$. This can be implemented by
\begin{flushleft}
\hspace{1mm}\includegraphics[scale= 0.73]{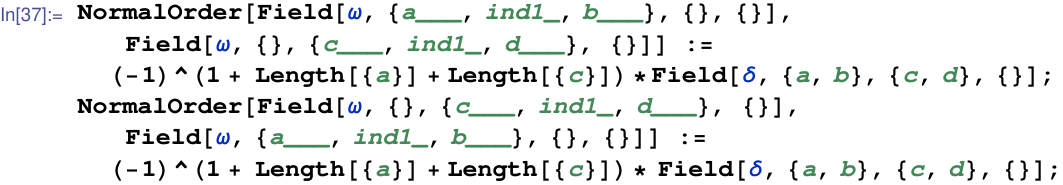}
\end{flushleft}
For simplicity, we choose $\omega$ to be constant, and write
\begin{flushleft}
\hspace{1mm}\includegraphics[scale= 0.73]{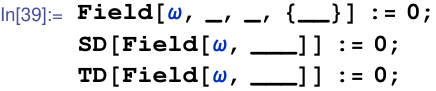}
\end{flushleft}
We now construct a composite even field $J$ by
\begin{flushleft}
\hspace{1mm}\includegraphics[scale= 0.73]{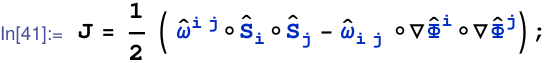}
\end{flushleft}
The operator $J$ is a primary field with conformal dimension $2$ with respect to $P$:
$\LB{P}{J}=(2 \partial + 2 \lambda + \chi D) J$, which is calculated by 
\begin{flushleft}
\includegraphics[scale= 0.73]{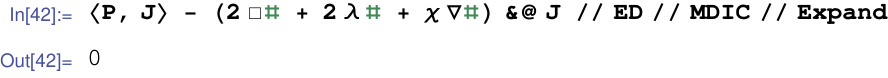}
\end{flushleft}
The algebra of $J$ with itself is calculated to $\LB{J}{J}=-( P + \text{dim} \lambda \chi)$:
\begin{flushleft}
\includegraphics[scale= 0.73]{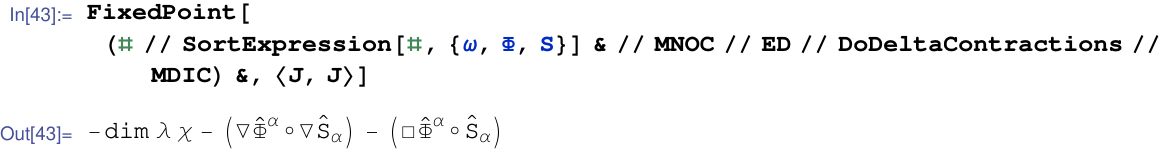}
\end{flushleft}
So, to conclude, $P$ and $J$ generates an $N=2$ superconformal algebra, with central charge  $c=3$ dim.




\section{List of commands} \label{sect:listofcommands}
In this section we list the commands provided by Lambda. 
\begin{description}
\item 
\cmd{CollectDerivatives[expr]}  tries to use Leibniz rule to write expressions 
involving derivatives more compact. Only works on the top-level of an 
expression. Short name is \cmd{CD[expr]}.
\item
\cmd{DefineComponents[s\_Symbol,c\_Symbol,t\_Symbol]} defines the expansion of the superfield with symbol \cmd{s} as $\mathtt{Field[c,\ldots] + \theta \;Field[t,\ldots]}$.
\item
\cmd{DoDeltaContractions[expr]} performs the possible contractions with the special field $\mathtt{\hat{\delta}_\alpha^\beta}$.
\item
\cmd{ExpandDerivatives[expr]} uses Leibniz rule to expand derivatives. E.g.\ $\mathtt{\Box (\hat{a} \circ \hat{b})}$ is
expanded to $\mathtt{(\Box  \hat{a} \circ \hat{b}) + ( \hat{a} \circ \Box \hat{b})}$ .
Short name is \cmd{ED[expr]}.
\item
\cmd{ExpandDummyIndices[expr, indRange\_List]} replaces all dummy indices with values from \cmd{indRange}, \eg, \cmd{ExpandDummyIndices[$\mathtt{\hat{a}_i \circ \hat{b}^i}$, \{1,2 \} ] } gives  $\mathtt{\hat{a}_1 \circ \hat{b}^1} +  \mathtt{\hat{a}_2 \circ \hat{b}^2}$. This is useful when given concrete realizations of tensors.
\item
\cmd{Field[sym\_Symbol,lind\_List,uind\_List,dind\_List]} is 
a field with symbol \cmd{sym}, lower indices are given by \cmd{lind}, 
upper indices by \cmd{uind} and the derivatives by \cmd{dind}. A 
symbol with a hat,\cmd{\^},  is interpreted as a field, and the 
indices are placed as, \eg, $\mathtt{\hat{a}^{u_1 u_2}_{l_1 l_2 , d_1 d_2}}$, with a comma separating the lower indices from the derivative indices.
\item
\cmd{FieldAntiSymmetricLowerIndices[sym\_Symbol]} declares that the 
field with the symbol \cmd{sym} is anti-symmetric in its lower 
indices.
\item
\cmd{FieldAntiSymmetricUpperIndices[sym\_Symbol] } declares that the 
field with the symbol \cmd{sym} is anti-symmetric in its upper 
indices.
\item
\cmd{FieldSymmetricLowerIndices[sym\_Symbol]} declares that the field 
with the symbol \cmd{sym} is symmetric in its lower indices.
\item
\cmd{FieldSymmetricUpperIndices[sym\_Symbol]} declares that the field 
with the symbol \cmd{sym} is symmetric in its upper indices.
\item
\cmd{GetSUSY[]} returns true, if the $\Lambda$-brackets are to be calculated in 
$N=1$ formalism, and false otherwise.
\item
\cmd{Grading[a]} returns an even value if \cmd{a} is regarded as an even (bosonic) field, and an odd value if \cmd{a} is regarded odd (fermionic).
\item
\cmd{LambdaBracket[a,b]} calculates the  $\Lambda$-bracket between \cmd{a} and \cmd{b}, \ie, $\LB{a}{b}$.
\item
\cmd{LambdaBracketChangeOrder[a,b]} calculates $\LB{a}{b}$ from the bracket $\LB{b}{a}$ using skew-symmetry.
\item
\cmd{LambdaBracketHandleIndices[a,b]} calculates $\LB{a}{b}$. First, the dummy indices are replaced with unique names, using \cmd{Make\-DummyIndices\-Unique}.  
After the calculation, the indices are made canonical, using \cmd{Make\-DummyIndices\-Canonical}. Inputed with the \cmd{AngleBracket}, i.e. $\mathtt{\langle a , b \rangle}$.
\item
\cmd{LambdaBracketJacobiator[a,b,c]}  calculates the jacobiator of the 
$\Lambda$-bracket, defined as 
\begin{equation*}
\LB{a}{\LB[\Gamma]{b}{c}} + (-1)^{1+N(a+1)} \LB[\Gamma + \Lambda]{ \LB{a}{b}  }{ c} +
                (-1)^{1+(a+N)(b+N)} \LB[\Gamma]{b}{ \LB{a}{c} }    
\end{equation*}
, see eq.\ \eqref{eq:Jacobiid}.
\item
\cmd{MakeDummyIndicesCanonical[expr]} renames repeated indices by the naming scheme $\alpha, \beta, \gamma$ etc. Greek letters should therefore be avoided as names for free indices.
Short name is \cmd{MDIC}.
\item
\cmd{MakeDummyIndicesUnique[expr]} makes repeated indices unique. Short name is \cmd{MDIU}.
\item
\cmd{MakeNormalOrderingCanonical[expr]} changes the normal ordering to the scheme $(\mathtt{a} \circ (\mathtt{b} \circ ( \mathtt{c} \circ \ldots )))$. Short name is \cmd{MNOC}.
\item
\cmd{MoveTerm[expr,pos,m]} moves the element at position \cmd{pos} of the expression \cmd{expr} \cmd{m} steps.
\item
\cmd{NormalOrder[a,b]} represents the normal ordered product of the fields \cmd{a} and \cmd{b}. It is denoted $\mathtt{(a \circ b)}$, \ie,   using the character \cmd{SmallCircle}. Write the input as \cmd{a} \textit{Esc} \cmd{sc} \textit{Esc} \cmd{b}. If more terms are given, the normal ordering binds from the right, so $\mathtt{a \circ b \circ c}$ is interpreted as   $\mathtt{(a \circ b) \circ c}$.
\item
\cmd{NormalOrderAssociator[a,b,c]} calculates $\mathtt{((a \circ b) \circ c) - (a \circ (b \circ c))}$ .
\item
\cmd{NormalOrderChangeOrder[a,b]} uses quasi-commutativity to express $(\mathtt{a} \circ \mathtt{b})$ as $(\mathtt{b} \circ \mathtt{a})+$ commutator. Short name is 
\cmd{NOCO}.
\item
\cmd{NormalOrderChangeParenthesis[expr]} uses \cmd{NormalOrderAssociator} to 
express $((\mathtt{a} \circ \mathtt{b} ) \circ  \mathtt{c} )$ or $(\mathtt{a} \circ ( \mathtt{b}  \circ  \mathtt{c} ))$ by means of the other expression. The expression $\mathtt{(a \circ  b )  \circ  (c \circ d )}$ is transformed to $\mathtt{((a \circ  b )  \circ  c ) \circ d )}$. Short name \cmd{NOCP}.
\item
\cmd{NormalOrderChangeParenthesisForward[expr]} transforms an expression of the form $\mathtt{(a \circ  b )  \circ  (c \circ d )}$ to the form $\mathtt{(a \circ (  b   \circ  (c  \circ d )))}$.
\item
\cmd{NormalOrderCommutator[a,b]} calculates  $(\mathtt{a} \circ  \mathtt{b} ) - (-1)^{|\mathtt{a}| |\mathtt{b}|}  (\mathtt{b} \circ  \mathtt{a} )$, where $\mathtt{|a|}$ denotes $\mathtt{Grading[a]}$.
\item
\cmd{OddNumericQ}  indicates that a variable is odd, \eg, like $\chi$. Note that $\mathtt{SD[\chi] = 2 \lambda}$ , but this is not
implemented for a general \cmd{OddNumericQ}.
\item
\cmd{SD[expr]} is the odd derivative on \cmd{expr}. Denoted $\mathtt{\nabla expr}$, defining the command \cmd{Del[expr]}. 
\item
\cmd{SetSUSY[True]} sets the $\Lambda$-bracket to be calculated within the $N=1$ formalisms. With \cmd{SetSUSY[False]}, $\Lambda$-brackets are calculated without superfields.
\item
\cmd{SortDerivativesFirst = False|True} sets, in case two fields have the same symbol, whether terms with derivatives should be sorted first or not.
\item 
\cmd{SortExpression[expr, order\_List] }  sorts the expression \cmd{expr}, in the order specified by \cmd{order}.
The list \cmd{order} has the symbols of the fields as values. E\. g., \cmd{SortExpression[$\mathtt{\hat a \circ \hat b}$], \{b,a\}]} yields \cmd{$\mathtt{\hat b \circ \hat a}$} + terms from 
moving the fields. If the symbols of the fields are not included in order, alphabetic order is used. If 
\cmd{SortDerivativesFirst} is set to true, terms with derivatives are sorted first, in case of equal symbols, otherwise last.
\item
\cmd{TD[expr]} is the even derivative on \cmd{expr}. Denoted $\Box \mathtt{expr}$, defining the command \cmd{Square[ expr ]}.
\end{description}

\section*{Acknowledgments}
I would like to thank U.\ Gran for inspiration to use Mathematica, see \cite{Gran:2001yh}. 
Moreover, I am grateful to  \mbox{R.\ Heluani}, \mbox{J.\ K\"all\'en} and \mbox{M.\ Zabzine} for many interesting and helpful discussions.  \mbox{J.\ K\"all\'en} also helped testing the package.
I am also thankful to  O.\ Ohlsson Sax, for discussions of possibilities and shortcomings of Mathematica.

\bibliographystyle{utphys}
\bibliography{lambdareferences}
 \end{document}